# INTEGRATING EXTERNALLY DEVELOPED SYSTEMS FOR SNS LINAC COOLING AND VACUUM

P.S. Marroquin, Los Alamos National Laboratory, Los Alamos, NM 87545, USA

## Abstract

External contractors are developing the local cooling and vacuum control systems for the Spallation Neutron Source (SNS) linac. Soon these systems will be integrated into the facility-wide controls system. Allen-Bradley Logix5000 series programmable controllers, populated with appropriate input/output modules, were selected as the local controllers. These controllers will be interfaced to the facility-wide control system via VME systems with PowerPC processors running the Wind River VxWorks operating system and Experimental Physics and Industrial Control System (EPICS) front-end controller software. This paper describes the interface and integration issues driven by project, cooling system and vacuum system requirements and hardware selections.

## 1 INTRODUCTION

Through much of the SNS, programmable logic controllers (PLC) will be used for the local controller systems. The EPICS input-output controllers (IOC) provide supervisory control, alarm functions and serve as the interface point for the PLC to global controls, the machine protection system and the event link system.

Integration of the local control systems to EPICS require resolution of interface issues preferably during design but in some cases issues are discovered during development and integration.

This paper examines a few of the more significant issues identified during design and development of the drift tube linac (DTL) and coupled cavity linac (CCL) resonance control and cooling system (RCCS) and vacuum systems. The issues examined were:

***Interface between IOC and PLC***; selection of a communication protocol, creating and reaching agreement on signal lists, mapping process variables (PV) to tags, IOC issuing commands in the form of requests, creating and naming tags for transfer to IOC, optimizing data transfer between PLC and IOC.

***Serial based devices***; use of serial ports on PLC and/or IOC, device driver development, serial types and network layout, distributed control.

***Control loops over Ethernet***; PID loops dispersed over IOCs and PLCs on non-dedicated networks.

***EPICS operator displays and controls vs. PLC displays***; cost, development and maintenance of local PanelView displays, redundant control screens on different platforms.

***Alarm checking and management***; the IOC as the appropriate location for alarm checking functions versus the PLC, alarm limit parameter maintenance.

More discussion of each of these interface and integration issues follow but first a brief background of the SNS DTL and CCL RCCS and Vacuum systems.

## 2 BRIEF SYSTEM DESCRIPTION

The focus of this paper is on the RCCS and vacuum systems of the SNS DTL and CCL linacs.

The RCCS (see Fig. 1) keeps the DTL tanks and CCL cavities in resonance by removing the RF waste heat from the copper cavity structures.

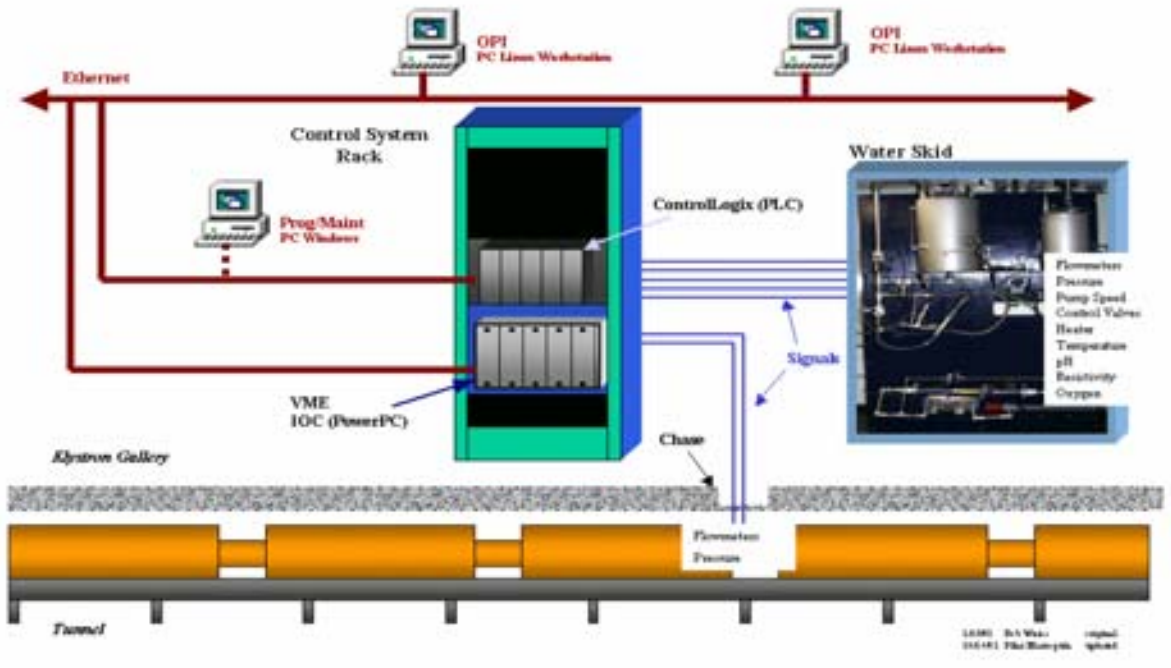

Figure 1: Resonance Control and Cooling System

Each DTL tank and CCL module has been designed to have its own vacuum control system, see Fig. 2.

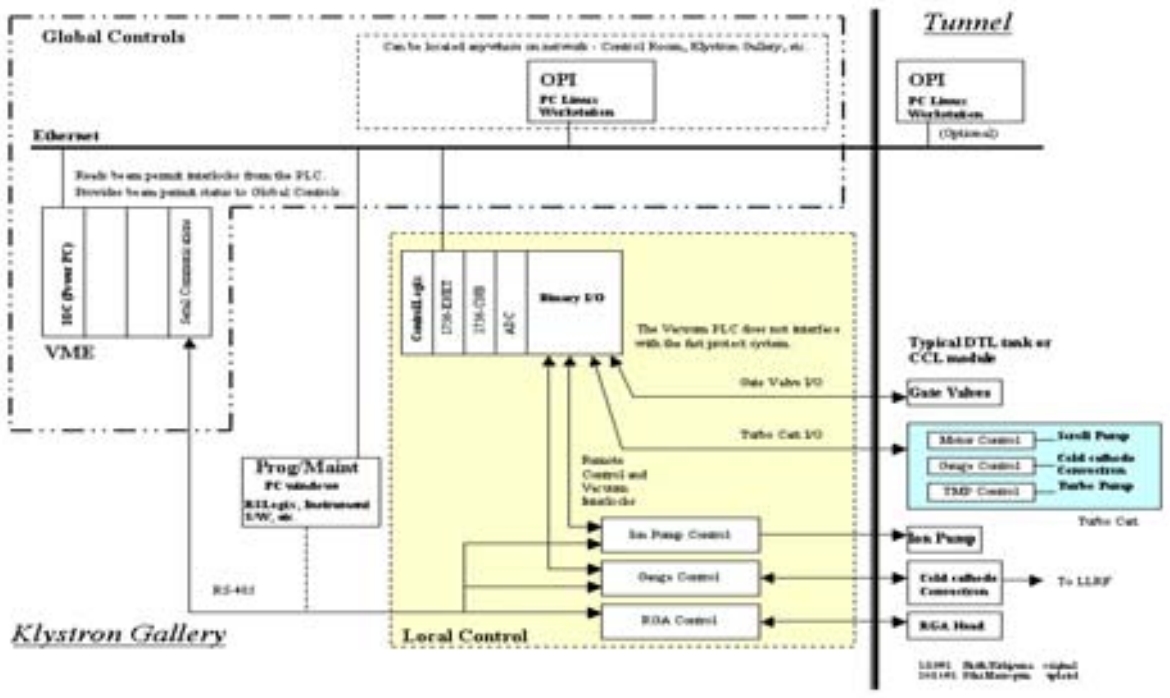

Figure 2: Vacuum System

The Allen Bradley ControlLogix programmable controllers are programmed with the Rockwell RSLogix5000 ladder code programming toolkit.

A network view of the RCCS and Vacuum System IOCs and PLCs are depicted in Fig. 3.

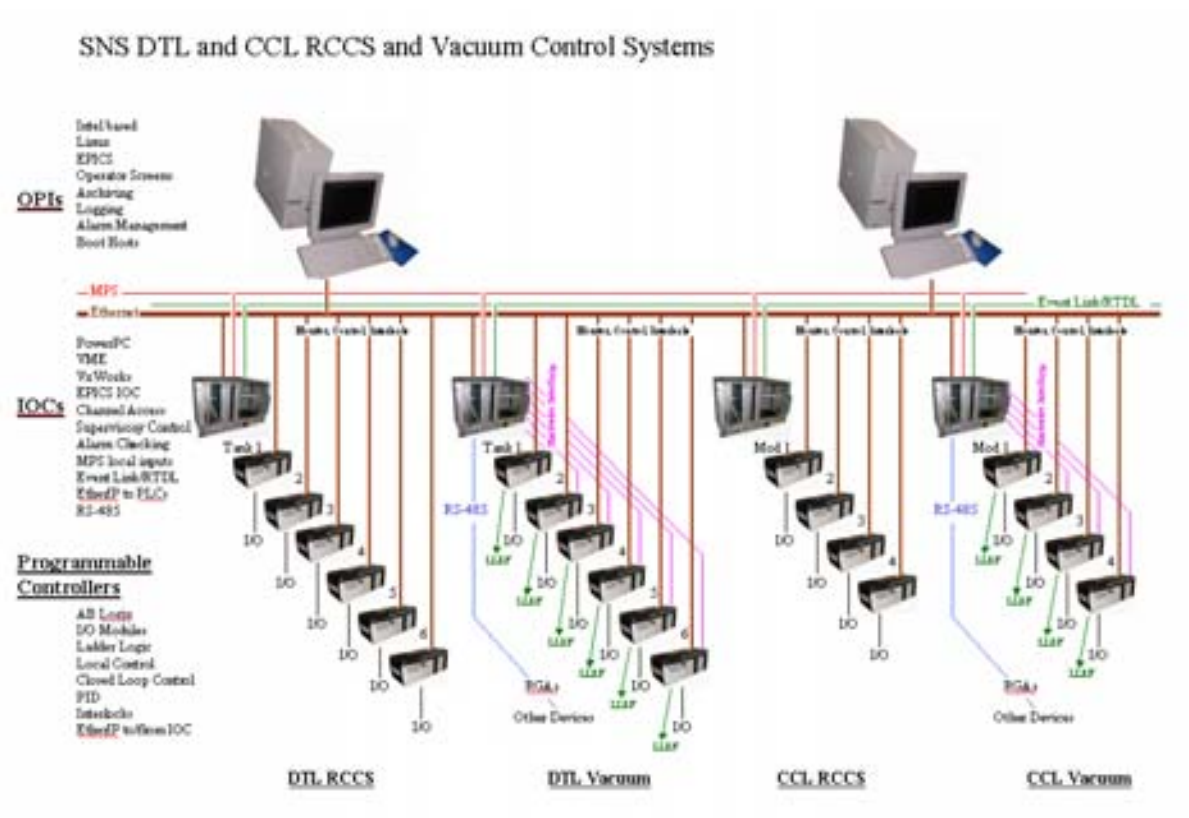

Figure 3: Network diagram of RCCS and Vacuum components

# 3 INTERFACE AND INTEGRATION ISSUES

## 3.1 Interface Between IOC and PLC

Two Control network options were under consideration, ControlNet and EtherNet/IP protocols.

ControlNet is an open real-time deterministic control network standard. The EtherNet/IP (Industrial Protocol) is an adaptation of ControlNet onto TCP/IP Ethernet.

The EtherNet/IP network was selected. The SNS Linac Controls group developed the device driver to use this technology in the IOC.

The drawbacks to this configuration stem from the foundations of Ethernet; a multi-node, peer-to-peer, non-deterministic, collision detection and back-off scheme network. This is not exactly suitable for real-time control, but it is acceptable when the traffic on a segment is limited and well known and the loop constants are slow.

The problems with ControlNet were that there is only one manufacturer, the interface board required programming which would result in programming the PLC, interface board and IOC all with the same information. The programming of the ControlNet board also requires a PC running windows which would add yet another architecture to support.

ControlLogix manages memory locations as named variables called tags allowing for arrays and mixed type data structures as well. Optimum tag-to-PV transfers require packing data into 500 byte arrays as often as possible. *(see ICALEPCS 2001 paper THAP020, Interfacing the ControlLogix PLC on EtherNet/IP, K.U. Kasemir, L.R. Dalesio).*

PLC handling of binary command values from the IOC is another issue. The PLC receives IOC commands as pulsed inputs to be latched and tested as ladder logic inputs.

## 3.2 Serial Based Devices

The requirements and design of the vacuum systems, market availability of specific types of instruments and collaboration with other SNS vacuum system design groups resulted in selection of instrument controllers with serial communications as the primary means for configuration, control and monitoring.

Although the goal was to install all I/O and control in the PLC, the IOC was chosen to host serial instruments. Serial ControlLogix modules, now available, involve IOC and PLC software development for full functionality. As the serial devices are not needed to support interlocks and displays are not needed at the PLC level, the IOC was a better choice.

Device driver and driver support development is required for housing the serial ports in the IOC. Serial device support within EPICS is abundant and more of a known quantity.

The control logic within the PLC may require inputs from the serial devices as well as initiate control but loop time requirements are in the order of seconds for which network latency would not cause problems.

Loss of IOC-PLC communication requires PLC ladder logic to be designed to handle such an event.

Serial network layout, specifically RS-485, is currently in the design stages. The physical locations of device controllers, performance requirements, expected network load per device and related subsystem functions are important considerations when laying out the RS-485 networks.

## 3.3 Control Loops over Ethernet

In the RCCS, it may be necessary to distribute the PID control loop across the IOC and the PLC. Mode change from RF control to temperature control also swaps out the measured variable and PID parameters.

Standard PID control software typically doesn't handle this correctly. The PID software in the IOC does. The configuration under consideration is for inputs into the PLC, data transferred to the IOC, PID in the IOC, control variables to the PLC, the PLC verifies control values and issues control signals to the final control elements.

Several issues arise from this configuration. One of primary concern is the loop time versus network latency. The cooling system is considered a slow process, in the order of seconds. The network latency has not been quantified to date but is considered a low concern. Total latency includes PLC scan time at both ends of the cycle and data transfer to and from and IOC scan time.

Loss of communication between the IOC and PLC must be monitored, on both controllers, with heartbeat mechanisms and contingency logic in place to handle such occurrences.

Loading of the IOC is also an issue. In the case of the DTL RCCS systems, one IOC supports the six PLCs. The IOC would be executing six independent PID loops along with the various other IOC functions (i.e. ~1000 PVs, Channel Access, record scanning, etc.).

### *3.4 Operator Displays*

The design called for each RCCS and Vacuum control rack to contain an Allen-Bradley PanelView display. The PanelView is an operator interface Windows CE based computer with a flat panel touchscreen monitor, all in one rack-mountable unit. Cost for the PanelViews range from ~$4000 for the 1000 model to ~$7000 for the 1000E model (graphics capability). The PanelView is configured from a Windows computer with specific Allen-Bradley software. Only ControlNet was available until recently, Ethernet is now available.

A cost cutting assessment of the design reduced the number of PanelViews by half. Later all of the PanelViews were eliminated after subsystem requirements versus display capabilities of EPICS were examined.

Development and maintenance costs of the displays were considered as well efforts being expended on developing and maintaining similar and redundant displays on the EPICS systems.

Early concerns over EPICS components not being ready for first installations of the system were the primary driver that added the PanelViews to the design. The EPICS IOC, OPI, channel databases and operator screens were developed and demonstrated successfully. EPICS IOCs and OPIs will be installed starting with first delivery.

### *3.5 Alarm Checking and Management*

To avoid duplication of alarm limit parameter sets and alarm checking, the IOC was chosen as the location for this function. The PLC will do limit checking if directly related to the controlled process and for equipment protection.

PLC data is sent to the IOC where they are alarm checked and flagged if determined to be in an alarm condition. EPICS clients detect the alarm conditions and react accordingly. The EPICS alarm manager centralizes management of alarms across the facility.

## 4 CONCLUSION

This paper described several significant interface and integration issues that were addressed during design and integration of the SNS DTL and CCL RCCS and Vacuum Systems to the facility-wide control system. The criteria followed to make these decisions include meet system requirements, provide reliable operation, minimize equipment cost, minimize development and maintenance efforts. By working closely with external systems integrators we are able to consider these issues from an integrated system perspective.

*Work supported by the Office of Basic Energy Science, Office of Science of the US Department of Energy, and by Oak Ridge National Laboratory.*